# Objective metrics for language lateralization of fMRI examinations: a new model for the classification of hemispheric dominance in healthy subjects and epileptic patients

**Authors:** M. Stroppi[1], D. Lizio[1], L. Berta[1], A. Citterio[2], C. Regna-Gladin[2], M. Rizzi[3], I. Sartori[3], P. E. Colombo[1], P. Arosio[4], A. Torresin[1]

[1] Department of Medical Physics, ASST Grande Ospedale Metropolitano Niguarda, Piazza Ospedale Maggiore 3, 20162 Milan, Italy;

[2] Department of Neuroradiology, ASST Grande Ospedale Metropolitano Niguarda, Piazza Ospedale Maggiore 3, 20162 Milan, Italy;

[3] Claudio Munari Epilepsy Surgery Center, ASST Grande Ospedale Metropolitano Niguarda, Piazza Ospedale Maggiore 3, 20162 Milan, Italy;

[4] Department of Physics, Università degli Studi di Milano, via Giovanni Celoria 16, 20133 Milan, Italy;

**Corresponding author:**
luca.berta@ospedaleniguarda.it

**Abstract**

**Purpose**: to compare different methods to calculate Laterality Index (LI), a metric which allows to evaluate hemispheric brain language dominance in functional MRI examinations (fMRI).

**Methods**: Two methods were considered for calculating LI: $LI_{AVE}$ and $LI_{VOL}$, respectively based on the differences between measurements of average and volume of fMRI signal in brain hemispheres. Laterality curves were obtained calculating values of $LI_{VOL}$ with increasing thresholds of fMRI signal and fitted with sigmoidal functions. A model for dominant and co-dominant classification based on fit parameters has been developed. The two methods and the sigmoidal model were applied to two cohorts of 93 epileptic patients and 27 healthy subjects undergoing language fMRI examinations with association, understanding and fluency tasks.

**Results**: Despite the different definitions, $LI_{AVE}$ and $LI_{VOL}$ resulted in equivalent classification of language lateralization. The agreement of neuroradiological clinical reports with classification of language lateralization resulting from the proposed methods ranged from 94.6% to 89.2% for LI metrics and up to 100% for the sigmoidal model. The fit parameters of the sigmoidal function defined empirical thresholds useful for classification between dominant and co-dominant, providing similar values for subjects and epileptic patients for fluency and association tasks. This result supports the idea of a unique model for language lateralization classification in epileptic patients and healthy subjects.

**Conclusions**: Language lateralization in fMRI can be effectively assessed by objective metrics. A novel approach based on sigmoidal fit of laterality curves resulted in higher agreement with clinical reports providing further information about the strength of language lateralization.

# 1. Introduction

Since the pioneering studies of Paul Broca, it has been recognized that the left cerebral hemisphere plays a fundamental role in language function [1]. This intuition, based on clinical and pathological observations, was later confirmed by results obtained with modern noninvasive neuroimaging techniques, such as functional magnetic imaging (fMRI) and tractography. These techniques also provided new information about the language network. Some neuroimaging studies, here below reported, hypothesized that from the earliest years of life, language is processed predominantly in the left cerebral hemisphere.

Holland et al. (2001) [2] studied a group of healthy children between the ages of 7 and 18 years and found an increase in left lateralization during these ages, and Szaflarski, Holland, Schmithorst, and Byars (2006) [3] found a statistically significant positive correlation between children's age (5-17 years) and lateralization indices for Broca's area. Wingfield [4] noted that language lateralization exhibits some changes during senescence, in fact there is a greater activation of the right hemisphere during language comprehension and production tasks among older subjects. This observation suggests that the degree of language lateralization decreases with age and cognitive processes become more symmetrical between the two hemispheres. In addition, transfer of language functions from the left to the right hemisphere is possible as an adaptive response to a pathology, such as a stroke or tumor in the left hemisphere. As it is evident from the examples just mentioned, it is not obvious to identify the areas involved in language functions, which must be determined precisely in the planning phase of a neurosurgical intervention, to minimize the risk of postoperative deficit (aphasia). Some studies showed that in healthy strongly right- and left-handed subjects language is predominantly lateralized in the left hemisphere, that right brain dominance rarely occurs, but above all that there is also a "bilateral representation of language", i.e. a condition of hemispheric co-dominance is present. This has been analyzed with modern neuro-imaging techniques, including PET, fMRI, tractography, and magneto-encephalography. For example, some fMRI studies [1] and [5] showed that tumors and epilepsy correlate with a more bilateral or right-handed representation of language, as pathology induces changes in the organization of the language network. Additionally, it was observed that although most areas of activation during language tasks are located in the left hemisphere, bilateral areas of activation are present when the right hemisphere is also involved.

However, defining which hemisphere is dominant is not so easy, since lateralization is not uniform across the different neuroanatomical components of the language network. Many studies showed that this is true in the case of healthy control subjects and even more in the case of patients with neurological diseases, such as focal epilepsy, where greater heterogeneities in language activation patterns are present (Tailbya et al [6]).

The review of Bernal and Ardila [1] reported many studies where it was showed that measures of hemispheric dominance, assessed with fMRI, are concordant with those of other techniques, including the Wada clinical test, functional transcranial Doppler ultrasound, and neuropsychological tests. Therefore, in recent years fMRI has become a fundamental tool in the clinical setting for the assessment of hemispheric linguistic dominance, especially since it is not an invasive technique, unlike other tests such as the Wada test. A synthetic index, called laterality index (LI), has been introduced in fMRI to quantify hemispheric dominance and facilitate its description. However the LI measure depends on several methodological factors, which should be controlled to assure that the LI results are meaningful. Primarily these factors include: location of the volumes of interest within each hemisphere (ROI), dependence on statistical threshold, threshold values of LI to establish hemispheric dominance, quantification of left and right hemisphere contributions (QLH and QRH), choice of tasks used, choice of baseline conditions, and reproducibility of LI values. Without accounting for these arbitrary factors, the LI measure could be seriously misleading.

The value of LI is calculated using the following standard formula [7]:

$$LI = f \cdot \frac{QLH - QRH}{QLH + QRH} \tag{1}$$

where QLH and QRH represent brain activity measured, during an fMRI examination, within the left and right hemispheres, respectively. Therefore a measure of LI summarizes, with a single value, the relative magnitude of brain activation of a given individual. The factor f defines the range of LI values, varying continuously from -f (right hemispheric dominance) to +f (left hemispheric dominance). Usually f has the value 1, so LI varies between -1 and +1. It should be noted that formula (1) was defined assuming that all quantities are positive, i.e. that QLH ≥ 0, QRH ≥ 0, QLH + QRH > 0. In addition, it is interesting to examine how this formula reflects the differences between left and right hemisphere contributions (QLH and QRH). Specifically, defining the relative difference R ∈ [-1, +∞[ between QLH and QRH as:

$$R = \frac{QLH - QRH}{QRH} \tag{2}$$

it is possible to rewrite the standard formula (1) as:

$$LI = f \cdot \frac{R}{R+2} \tag{3}$$

According to Seghier [7], if the absolute value of LI value is lower than 0.2 a patient/subject is classified as co-dominant otherwise he is classified as dominant. This value of 0.2 was suggested for healthy subjects, instead for epileptic patients it was argued that a threshold value could be higher due to the different organization of the linguistic network [6].

The aim of this study was to compare different methods of attributing the laterality of language by creating a model based on fMRI data that can be implemented in every-day clinical practice. In particular the LI was calculated by evaluating the activations obtained in different brain regions (ROIs) for different fMRI tasks (association, comprehension and fluency) administered to patients with drug-resistant epilepsy and healthy subjects with two different methods based on average and volumes of intensity of the 3D activation pattern distributions. Moreover, by studying the trend of LI values as a function of the applied threshold, a model based on sigmoidal fit was developed in order to classify the hemispheric dominance or co-dominance of the language function for each task.

## 2. Material and Methods

All data were acquired as part of the Centre's Presurgical evaluation programme, and data were analysed retrospectively. All patients agreed by signing an appropriate form.

### 2.1. Partecipants

In this study laterality index of 93 epileptic patients, 50 males and 43 females, was evaluated, mean age 33.9 ± 14.8 years, undergoing treatment at "Claudio Munari" Epilepsy Surgery Center, ASST Grande Ospedale Metropolitano Niguarda. In addition, laterality index was calculated in 27 healthy subjects, 15 males and 12 females, mean age 36.3 ± 10.9 years. Patients and subjects were divided into two categories: dominant and co-dominant, according to "clinical reports of fMRI data" review by experienced neuroradiologist.

## 2.2. MRI Data acquisition

Data were collected on 1.5T Philips Achieva MR scanner. The experimental protocol first envisaged the acquisition of high spatial resolution T1-weighted anatomical volume, using a Fast Field Echo (FFE) sequence. The following acquisition parameter values were set: TR = 7.22 ms, TE = 3.24 ms, flip angle = 8°. For each scan 180 sagittal slices were acquired, with thickness of 0.9 mm, acquisition matrix of 256 × 256 pixels, pixel size equal to 0.46 mm. The anatomical volume was therefore representable through a 3D matrix with the following dimensions: 256 × 256 × 180. The single voxel element measured: 0.46 mm × 0.46 mm × 0.9 mm. fMRI data were obtained through $T_2^*$ EPI echo gradient sequences. Acquisition parameters were the following: TR = 3000 ms, TE = 50 ms, flip angle = 90°. For each scan 30 axial slices were acquired, with thickness of 4.4 mm, acquisition matrix of 128 × 128 pixels, pixel size equal to 1.8 mm.

| Protocol Name | T1 (FFE) | FMRI (G-EPI) |
|---|---|---|
| Plane | Sagittal | Axial |
| TE (ms) | 3.24 | 50 |
| TR (ms) | 7.22 | 3000 |
| FA (flip angle) | 8° | 90° |
| Acquisition Matrix | 256x256 | 128x128 |
| In plane pixel size (mm$^2$) | 0.46x0.46 | 1.8x1.8 |
| Slices | 180 | 30 |
| Slice thickness (mm) | 0.9 | 4.4 |
| Acquisition time (min) | 7 | 6 |

*Table 1: Acquisition parameter values.*

## 2.3. Experimental paradigm

Three tasks in separate sessions were used, each of which allowed to probe different aspects of language: association of names, understanding of sentences, phonemic and semantic fluency. Specifically, the experimental design was a block diagram of task/rest type, in which the stimuli were presented as an alternation of a task period, 30 s, and a rest one, 30 s. The train of blocks had a total duration of 6 minutes.

## 2.4. Pipeline for fMRI data processing

A pipeline was developed in Python language to automatically calculate LI values starting from images obtained in an fMRI exam. The pipeline is based on Nipype [8] (Neuroimaging in Python: Pipelines and Interfaces https://nipype.readthedocs.io), an open source software that enables to interface automatically with neuroimaging software's tools using a standard syntax in a Python program. Specifically was used FSL's tool FEAT [9] in order to analyze fMRI data (FSL: https://fsl.fmrib.ox.ac.uk/fsl/fslwiki [10], [11] and [12]). The LI was calculated from activation patterns in the regions of Broca (BA44 and BA45), Wernicke and hemisphere (respectively Figure 1a, 1b and 1c) as defined by the atlases Juelich and Harvard-Oxford Cortical Structural Atlas, as these brain regions are indicated in the literature as being related to language functions [13], [14] and [15]. At neurological level, during association and fluency tasks mainly Broca area is activated [13], [14] and [15], while during comprehension task mainly Wernicke area is activated [13], [14] and [15]. During comprehension task cognitive mechanisms can be influenced by auditory processes [16] that results in bilateral activations. LI was also calculated in hemisphere ROI in order to evaluate the totality of activations, even in areas not directly related to linguistic functions (such as motor area and Cerebellum). Using FSL's tools it is possible to apply to images statistical analysis that allows to evaluate which areas show activations significantly different from the background level and to obtain pattern intensity map. A "threshold"-value is set in order to discriminate between significant activations and background, i.e. set to zero voxel's values which signal is lower than threshold value. Two methods were implemented for calculating LI: $LI_{AVE}$ and $LI_{VOL}$ respectively based on the "average" and "volume" methods. $LI_{AVE}$ values were calculated from the average intensity of the pattern of activation resulting from fMRI analysis (Z-statistic image) in a ROI, applying a fixed threshold calculated by Gaussian Random Field (GRF) [17]. $LI_{VOL}$ values were calculated starting from the number of the activated voxel in a ROI, by applying to Z-statistic image a percentage threshold of 60% of the maximum intensity in the activation pattern of each task. The value of 60% was established after preliminary evaluations. LI values of 93 epileptic patients were evaluated with both methods and compared with clinical reports provided by neuroradiologists. In addition, LI values of 27 healthy subjects were calculated.

All images processing was done on computer HP Z 600- Intel(R)Xeon(R) X5650 2.67GHz, 12GB RAM, Ubuntu 20.04, 64bit.

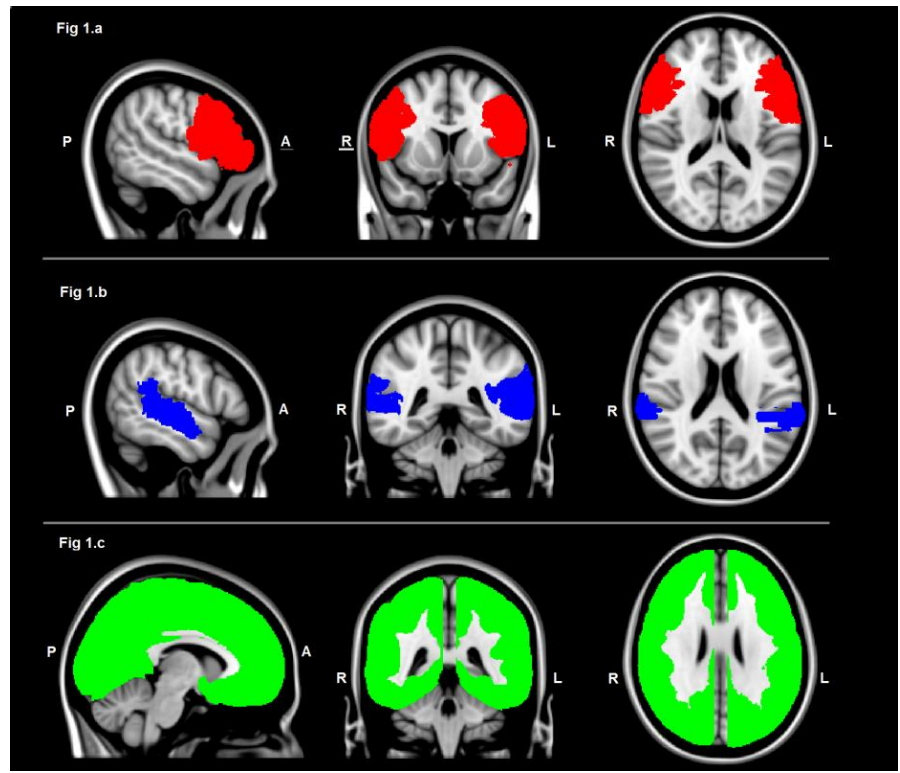

*Fig. 1: Figure 1a ROI of Broca's area (left), 1b ROI of Wernicke's area (center) and 1c ROI of hemisphere's area (right).*

## 2.5. Fitting laterality curves

Since standard expression (1) for LI is a ratio between the different patterns of activations in the two hemispheres, if a condition of hemispheric dominance is present, the trend of LI's value as function of a percentage threshold (volume method) should be a sigmoid. Specifically, laterality curves obtained calculating $LI_{VOL}$ with different threshold values were fitted with standard expression of a sigmoid:

$$f(x) = \frac{\pm 1}{1+\exp(-x \cdot a + b)} \qquad (4)$$

where "a" and "b" are two characteristic parameters of the fit. The sigmoid functions and the fitting parameters were calculated for all patients and subjects, for all tasks in ROI Specific and hemisphere ROI. Gnuplot software (http://www.gnuplot.info), based on the implementation of Marquardt-Levenberg algorithm of non-linear least squares (NLLS) [18], [19] was used for the best fitting.

## 3. Results

In Table 2 the number of patients and subjects classified as dominant or co-dominant is reported according to neuroradiologists clinical reports for the 3 functional tasks. Using the standard classification dominance/co-dominance, i.e. using a value of 0.2 as LI [7], the agreement between LI values and neuroradiologists clinical reports ranges is reported in Table 3. A perfect agreement is found for subjects while rates up to 16% of disagreement were found for epileptic patients. As expected, since the symmetrical auditory component is particularly involved during the comprehension task [16], the number of individuals with co-dominant activation in this task is greater than those in fluency and association tasks. This evidence is highlighted by neuroradiological reports and correctly provided by LI metrics. Using $LI_{AVE}$ metrics, an agreement of 94.6% and 89.2% was found when LI were calculated using Specific-ROI and hemispheric ROI, respectively. In particular, in n=5 cases no agreement was found using $LI_{AVE}$ calculated in the specific ROI or in the hemisphere. These patients, classified as co-dominant by clinical reports, resulted with LI values slightly higher than 0.2. Furthermore, in all cases where $LI_{AVE}$ calculated in the hemispheric ROI disagreed with clinical reports (n=5), the activation patterns were described as single and separated spots. Similar results were obtained using $LI_{VOL}$ metrics: the agreement with clinical reports were 89.2% and 83.8% for specific and hemispheric ROIs, respectively.

| **Association Task** | Total number | Co-dominant (Broca) | Dominant (Broca) | Co-dominant (hemisphere) | Dominant (hemisphere) |
|---|---|---|---|---|---|
| Subjects | 27 | 4 | 23 | 3 | 24 |
| Patients | 93 | 9 | 84 | 9 | 84 |
| **Fluency Task** | Total number | Co-dominant (Broca) | Dominant (Broca) | Co-dominant (hemisphere) | Dominant (hemisphere) |
| Subjects | 27 | 0 | 27 | 3 | 24 |
| Patients | 93 | 10 | 83 | 10 | 83 |
| **Understanding Task** | Total number | Co-dominant (Wernicke) | Dominant (Wernicke) | Co-dominant (hemisphere) | Dominant (hemisphere) |
| Subjects | 27 | 12 | 15 | 12 | 15 |
| Patients | 93 | 33 | 60 | 32 | 61 |

*Table 2: Number of patients and subjects classified as dominant or co-dominant according to neurological clinical reports, in specific ROI (Broca and Wernicke) and in hemisphere ROI for the three tasks.*

| | Number | $LI_{AVE}$ Specific-ROI | $LI_{AVE}$ hemisphere ROI | $LI_{VOL}$ Specific-ROI | $LI_{VOL}$ hemisphere ROI |
|---|---|---|---|---|---|
| Patients | 93 | 88 (94.6%) | 83 (89.2%) | 83 (89.2%) | 78 (83.8%) |
| Subjects | 27 | 27 (100%) | 27 (100%) | 27 (100%) | 27 (100%) |

*Table 3: Agreement between LI values and clinical reports in Specific ROI (Broca and Wernicke) and in hemisphere ROI, for patients and subjects.*

### 3.1. Laterality curves

$LI_{AVE}$ and $LI_{VOL}$ metrics summarize in a single value of simple interpretation all the information described by the distribution of activation patterns. While no other parameters are needed for $LI_{AVE}$ calculation, in the case of $LI_{VOL}$ a threshold level of 60% of maximum intensity is defined. A more effective approach to describe the complex patterns of activation and to quantify language lateralization might be represented by a metric that accounts for volumes of activation calculated at different thresholds values. For this purpose, lateralization curves are introduced. Examples of lateralization curves for epileptic patients having dominant and co-dominant language representation are reported in figure 2,3 and 4. In the graphs, the $LI_{AVE}$ values are also reported for comparison purposes. Considering $LI_{AVE}$ and $LI_{VOL}$ in hemisphere and in specific ROI, LI values in hemisphere ROI are lower than corresponding values in specific ROI. With hemisphere ROIs all activations are considered, also false positives or activation in other cortical areas not related to language functions. However, LI calculated in both specific and hemisphere ROI describes properly language lateralization.

*Fig. 2: Laterality curves for a left-dominant patient computed in specific ROI (Broca and Wernicke) and in hemisphere ROI for association, understanding and fluency task, respectively. With low thresholds values, $LI_{VOL}$ is lower than 0.2, indicating a bilateral representation of the language, while at higher threshold $LI_{VOL}$ increase up to 1, indicating left representation of language. $LI_{VOL}$ and $LI_{AVE}$ correctly describe language lateralization for a threshold of 60% of the maximum value of the task.*

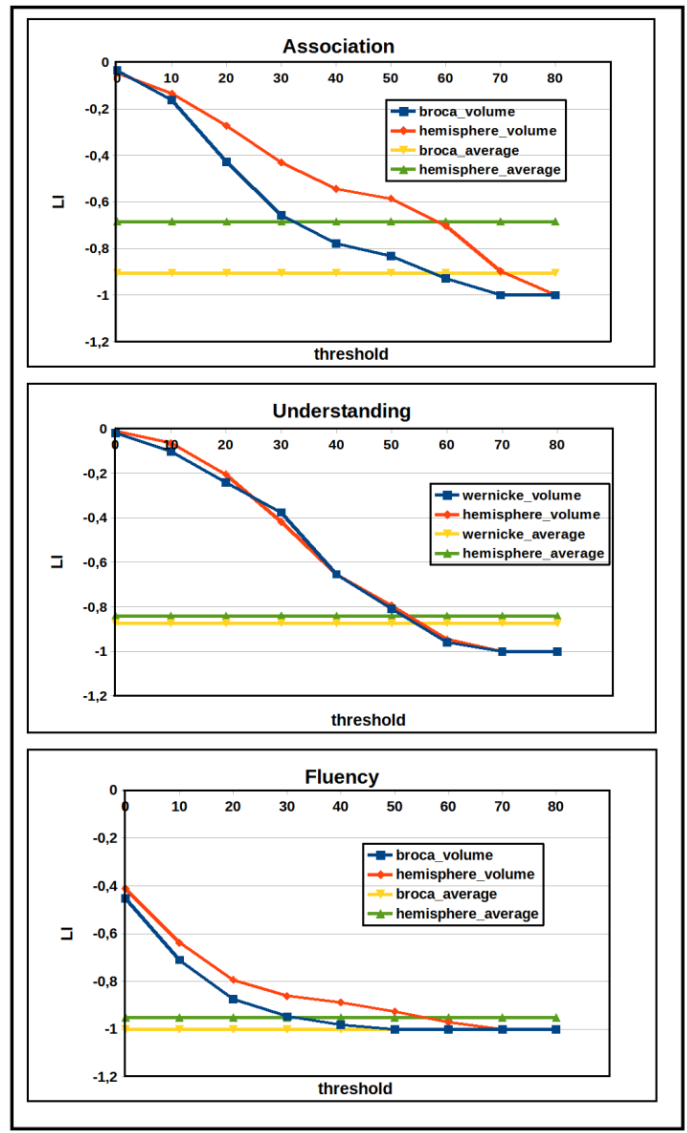

*Fig. 3: Laterality curves for a right-dominant patient computed in specific ROI (Broca and Wernicke) and in hemisphere ROI for association, understanding and fluency task, respectively. With low thresholds values, $LI_{VOL}$ is lower than 0.2, indicating a bilateral representation of the language while at higher threshold $LI_{VOL}$ decrease up to -1, indicating right representation of language. $LI_{VOL}$ and $LI_{AVE}$ correctly describe language lateralization for a threshold of 60 % of the maximum value of the task.*

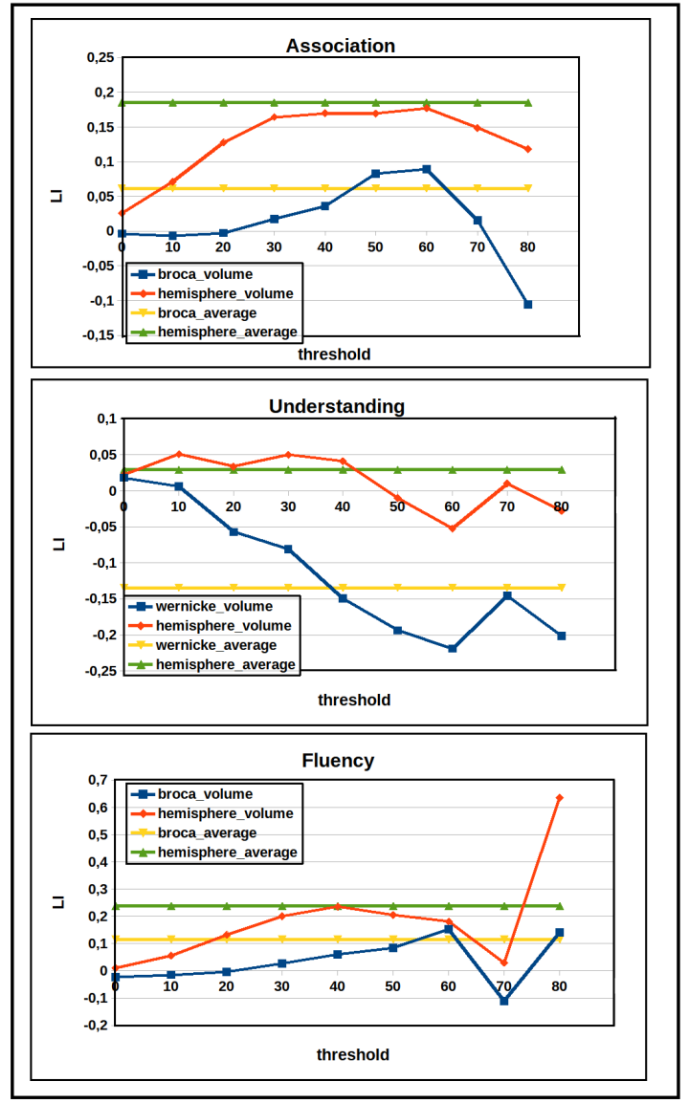

*Fig. 4: Laterality curves for a co-dominant patient computed in specific ROI (Broca and Wernicke) and in hemisphere ROI for association, understanding and fluency task, respectively. $LI_{VOL}$ and $LI_{AVE}$ correctly describe language lateralization for a threshold of 60 % of the maximum value of the task.*

### 3.2. Sigmoid model for lateralization

By comparing LI's trends as function of a percentage threshold showed in Figures 2, 3 and 4 for the three tasks, it is possible to observe that trends of left hemispheric dominance are specular to those of right hemispheric dominance and both are different from those obtained in individuals with bilateral language representation.

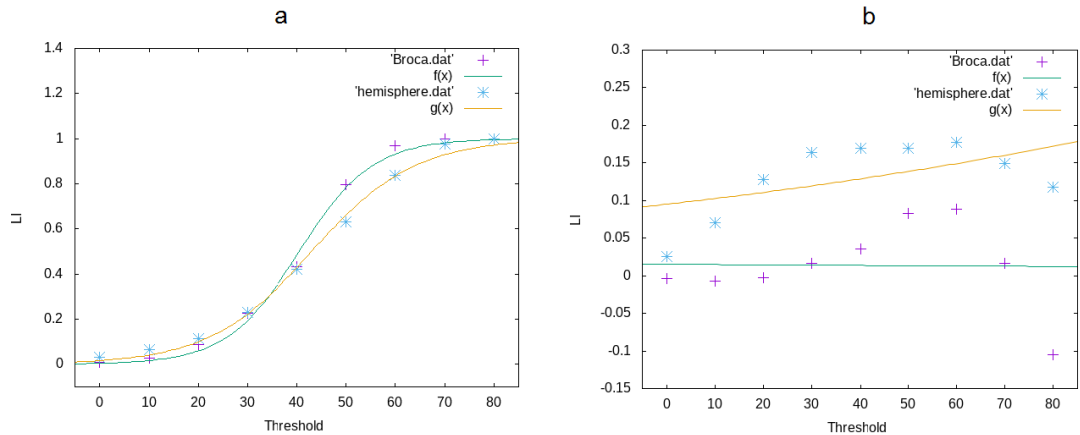

*Fig. 5: Laterality curves and sigmoidal Fits for dominance (left) and co-dominance (right) patients relative to specific ROI and hemisphere ROI. In case of language dominance the fit describes LI's trends, whereas in case of language co-dominance, the data are not well fitted resulting in low slope of the sigmoid function, described by "a" value.*

These different properties of the laterality curves are well described by the fit parameter "a". In particular, dominant and co-dominant individuals show different distributions of "a" values, according to T-tests. All results of statistical tests obtained for all functional tasks and different ROIs are shown in table 4.

Analyzing the values of the fit parameters "a" and clinical reports of subjects and patients, it is possible to define empirical thresholds useful for classification between dominant and co-dominant, as reported in Table 5.

The percentage agreement between language lateralization classified using such threshold values and clinical reports was 100% for association and fluency tasks in epileptic patients.

4. **Discussion**

In this work two methodological approaches for the calculation of the laterality index of language activation in fMRI tasks are evaluated. $LI_{AVE}$ and $LI_{VOL}$ calculated with a fixed threshold (60% of maximum of the intensity pattern) lead to equivalent conclusions regarding the language lateralization. An alternative approach based on laterality curves was proposed to overcome the limitations in this classification task based on a fixed threshold value of LI.

According to our results, a threshold value of 0.2 as LI effectively describe language lateralization in healthy subjects, whereas in case of epileptic patients it may be misleading. In fact, in our cohort 5 epileptic patients were classified as co-dominant by clinical report even LI values were higher than 0.2.

In these situations the Sigmoid Model, based on parameters "a", representing the slope of the LI values calculated with increasing thresholds of the activation pattern intensity, has been shown to be robust in classifying language lateralization of subjects and epileptic patients too. Since dominant and co-dominant individuals show different laterality curves, a correct classification based on "a" values resulted in all epileptic patients.

Figures 6, 7 and 8 show bi-dimensional plots of the sigmoid parameters "a" and "b" calculated for association, understanding and fluency tasks, respectively. Data of subjects and patients are reported for fit parameters obtained in specific and in hemisphere ROIs.

Comparing the resulting threshold values of "a" for differentiating dominant and co-dominant individuals reported in Table 5, similar values are found for subjects and epileptic patients for fluency and association tasks supporting the idea of a unique model for language lateralization.

Further information about lateralization could be deduced using the values of "b" parameters. The fit parameter "b", represents the threshold value at which the lateralization curve measure 0.5. Individuals with strong lateralization are associated with low value of "b" parameters and vice-versa. With this perspective, a further classification of language network is possible.

This work has some limitations. The classification between dominant and co-dominant individuals is based on clinical interpretation of the activation pattern of fMRI examinations. In the neuroradiologist's interpretation of the pattern of activations of fMRI data may be influenced by statistical noise, therefore the clinical interpretations may be challenging. Another independent test would be required for accurate diagnosis.

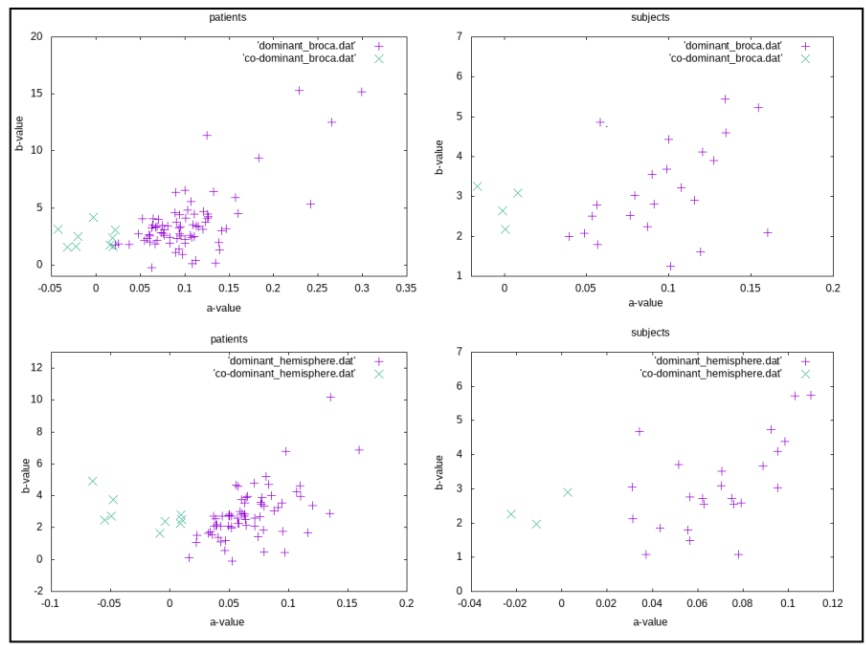

*Fig. 6: Fit parameter "b" as function of parameter "a" for association task. Dominant and co-dominant individuals show different distributions of "a" values, according to T-tests. Analyzing the values of the fit parameters "a" it is possible to define empirical thresholds useful for classification between dominant and co-dominant. The purple symbol indicates individuals classified (before the fit) by clinical reports as dominant, while the green symbol indicates individuals classified (before the fit) by clinical reports as co-*

*dominant. Left and right panel of image shows results obtained for patients and subjects, respectively. Upper and lower panel of image shows results obtained in specific ROI of Broca and in hemisphere ROI, respectively.*

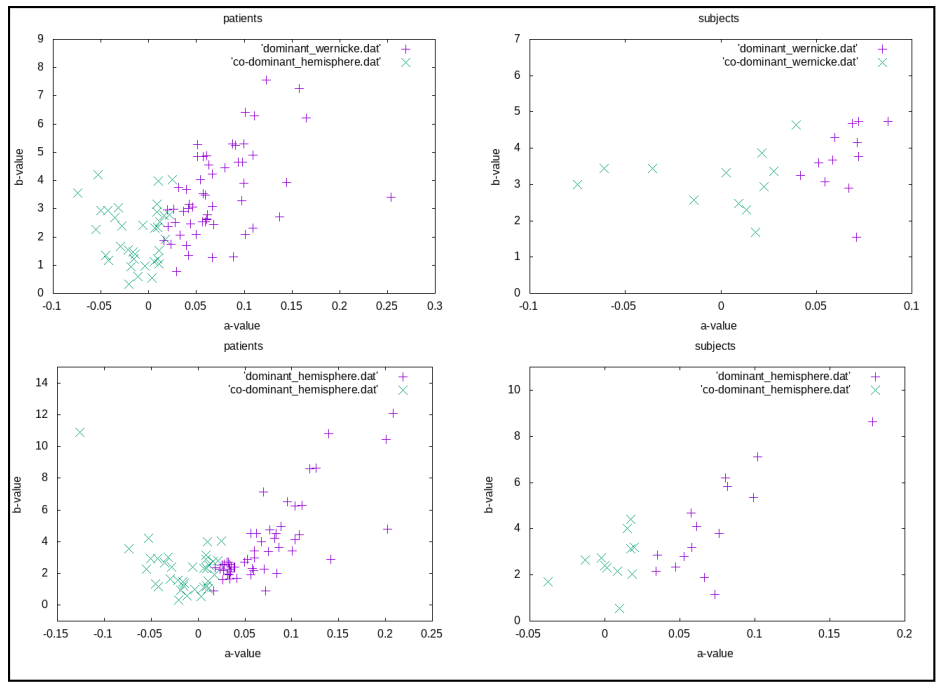

*Fig. 7: Fit parameter "b" as function of parameter "a" for comprehension task. Dominant and co-dominant individuals show different distributions of "a" values, according to T-tests. Analyzing the values of the fit parameters "a" it is possible to define empirical thresholds useful for classification between dominant and co-dominant. The purple symbol indicates individuals classified (before the fit) by clinical reports as dominant, while the green symbol indicates individuals classified (before the fit) by clinical reports as co-dominant. Left and right panel of image shows results obtained for patients and subjects, respectively. Upper and lower panel of image shows results obtained in specific ROI of Wernicke and in hemisphere ROI, respectively.*

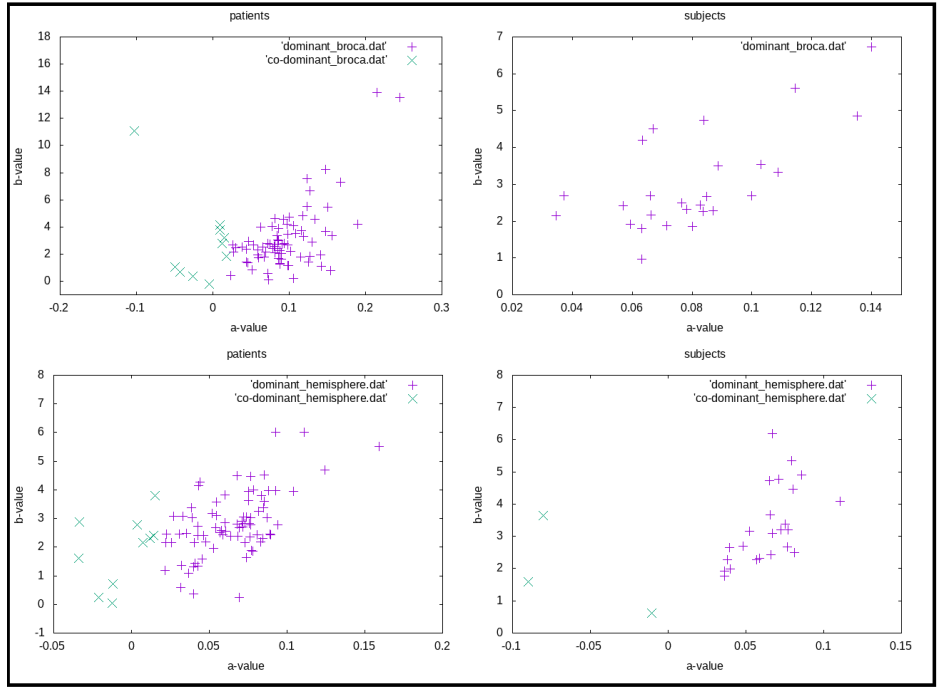

*Fig. 8: Fit parameter "b" as function of parameter "a" for fluency task. Dominant and co-dominant individuals show different distributions of "a" values, according to T-tests. Analyzing the values of the fit parameters "a" it is possible to define empirical thresholds useful for classification between dominant and co-dominant. The purple symbol indicates individuals classified (before the fit) by clinical reports as dominant, while the green symbol indicates individuals classified (before the fit) by clinical reports as co-dominant. Left and right panel of image shows results obtained for patients and subjects, respectively. Upper and lower panel of image shows results obtained in specific ROI of Broca and in hemisphere ROI, respectively.*

| **Association** | patients | subjects |
|---|---|---|
| Broca | $3.26 \times 10^{-9}$ | $2.40 \times 10^{-9}$ |
| hemisphere | $1.05 \times 10^{-5}$ | $6.88 \times 10^{-4}$ |
| **Understanding** | patients | subjects |
| Wernicke | $6.13 \times 10^{-16}$ | $1.30 \times 10^{-1}$ |
| hemisphere | $9.00 \times 10^{-15}$ | $1.28 \times 10^{-6}$ |
| **Fluency** | patients | subjects |
| Broca | $2.00 \times 10^{-8}$ | x |
| hemisphere | $1.46 \times 10^{-8}$ | $3.54 \times 10^{-2}$ |

*Table 4: t-test values. Using t-test it is possible to verify that in every panel of image the two datasets are statistically distinct (regarding parameter "a"). The × symbol indicates that it is not possible to obtain results from t-test because there are not enough data.*

| Association | Co-dominance (patients) | Dominance (patients) | Co-dominance (subjects) | Dominance (subjects) |
|---|---|---|---|---|
| Broca | <0.022 | >0.022 | <0.020 | >0.020 |
| hemisphere | <0.013 | >0.013 | <0.010 | >0.010 |
| **Understanding** | Co-dominance (patients) | dominance (patients) | Co-dominance (subjects) | Dominance (subjects) |
| Wernicke | <0.020 | >0.020 | <0.040 | >0.040 |
| hemisphere | <0.017 | >0.017 | <0.027 | >0.027 |
| **Fluency** | Co-dominance (patients) | dominance (patients) | Co-dominance (subjects) | Dominance (subjects) |
| Broca | <0.023 | >0.023 | X | X |
| hemisphere | <0.018 | >0.018 | <0.020 | >0.020 |

5.

*Table 5: Critical values of the parameter "a" which allow to discriminate between condition of hemispheric dominance and co-dominance of language.*

## 6. Conclusions

In this work a Python program was developed to calculate the LI from fMRI images, an objective metric to classify language lateralization in patients and subjects. The two methods for the calculation of LI, based on volume and average intensity of activation patterns, resulted equivalent but with the novel approach based on sigmoidal fit of laterality curves resulted in higher agreement with clinical reports providing further information about the strength of language lateralization. LI values and fit parameters of the sigmoidal fit can be reported in the clinical workflow of image analysis